\documentclass[aps,prb,twocolumn,floatfix,superscriptaddress,showpacs,amsmath,amssymb]{revtex4}
\usepackage{graphicx}
\usepackage{epsfig} 
\allowdisplaybreaks

\newcommand{\be}{\begin{equation}}
\newcommand{\ee}{\end{equation}}
\newcommand{\bea}{\begin{eqnarray}}
\newcommand{\eea}{\end{eqnarray}}
\newcommand{\ba}{\begin{eqnarray*}}
\newcommand{\ea}{\end{eqnarray*}}

\newcommand{\mR}{\mathcal{R}}
\newcommand{\mI}{\mathcal{I}}

\newcommand{\bk}{\mathbf{k}}

\newcommand{\m}[1]{\mathcal{#1}} 
\newcommand{\eps}{\varepsilon}

\begin{document}

\title{Nonequilibrium Dynamics Across an Impurity Quantum Critical Point}
\author{Marco Schir\'o}
\affiliation{Princeton
  Center for Theoretical Science and Department of Physics, Joseph Henry
  Laboratories, Princeton University, Princeton, NJ
  08544} 
\date{\today} 
 

\begin{abstract} 
Whether a small quantum mechanical system is able to equilibrate with its environment once an external local perturbation drives it out of thermal equilibrium is a central question which cuts across many different fields of science. Here we consider such a problem for a correlated quantum impurity coupled to a fermionic reservoir and driven out of equilibrium by \emph{local quantum quenches} such as those recently realized in optical absorption experiments on single quantum dots.
We argue that equilibration in this problem is deeply connected to the occurrence of Kondo Effect at low energy and that a highly non trivial dynamical behaviour may emerge whenever a local quantum critical point intrudes between a conventional Kondo screened phase and a Kondo unscreened one. We discuss this issue in the context of the Anderson Impurity model coupled to a pseudo-gap reservoir by using a correlated time dependent variational wave function that is able to qualitatively describe this physics

\end{abstract}
\maketitle

\textit{Introduction -} 
Small interacting quantum mechanical systems coupled to an external environment represent basic paradigms of transport, dissipation and non equilibrium phenomena. Understanding their dynamical behaviour and to what extent those system are able to equilibrate with their environment 
when a local perturbation drives them far from equilibrium is therefore crucial in many different physical contexts. While for systems coupled to a bosonic environment this question has been around since the early days of Caldeira-Leggett models~\cite{Caldeira_Leggett_PRL81,Leggett-RMP} and it is keep attracting a lot of interest~\cite{spin-boson,VonDelft_arxiv11,Kirchner_arxiv11},  much less is generally known about the out of equilibrium relaxation dynamics of small quantum systems coupled to fermionic reservoirs.
 Experimental realizations of these fermionic quantum impurity models traditionally involve diluted magnetic impurities in metals~\cite{Hewson_book}, quantum dots and single molecules attached to leads~\cite{Goldhaber_Gordon_nature98,Kouwenhoven_science98, GoldhaberGordon_prl08_noneq,Park_Coulomb_Blockade,Florens_C60} or magnetic adatoms on metallic surfaces~\cite{Madhavan24041998}. While well suited to study transport, these setup do not probe directly the nonequilibrium relaxation dynamics. Recent achievements in controlling optical properties of single semiconducting quantum dots brought a new twist in quantum impurity physics. By shining coherent light on those systems, experiments are able to probe the out of equilibrium dynamics induced by a sudden switching of a local perturbation, a so called \emph{local quantum quench}~\cite{Helmes_prb2005,Heyl_Kehrein_arxiv2010,Tureci_prl11,VonDelft_AOC_arxiv11,Andrei_PRB12} . The experimental evidence of the Kondo effect, a smoking gun of strong local correlations, in the absorption spectrum of a single self assembled quantum dot~\cite{Imamoglu_nature11} opens the exciting possibility to explore in a highly controlled setup the non equilibrium dynamics of quantum impurity models coupled to fermionic environments.
A fundamental question this experimental breakthrough brings over is whether and under which condition a quantum impurity is able to equilibrate with its fermionic environment after a local perturbation, such as a sudden switch-on of the coupling with the bath, drives it far from equilibrium.
A very well studied example along this line is the Anderson Impurity Model (AIM)~\cite{AIM_meanfield} a paradigmatic model describing a single interacting electronic level coupled to a fermionic reservoir. A number of theoretical investigations on its out of equilibrium dynamics found that quite generically thermalization after a local quantum quench occurs~\cite{Nordlander_99,Kehrein_Lobaskin_05,tnrg_Anders,Anders_prb06} . Yet,  when the system is deep in the Kondo regime an exponentially long  time scale controls this relaxation, as the system flows
toward the strong coupling (SC) fixed point.
The Kondo effect is, however, not the only possible fate for a quantum impurity at low energy.
A different and perhaps more intriguing possibility, which motivates the present paper, may arise whenever a mechanism competing with the Kondo effect is present, which is able to drive the system away from the SC fixed point and eventually to renormalize to zero the effective coupling with the bath. This is what happens when an impurity quantum critical point intrudes between a SC Kondo screened phase and a Local Moment (LM) phase. Here we study the out of equilibrium real time dynamics across such an impurity quantum criticality. We consider the simplest example showing this competition, namely the Anderson impurity model coupled to a pseudogap reservoir~\cite{Fradkin_PRL,Ingersent_PRB,FritzVojta_prb04}. By using a time dependent correlated variational wave function we show that an highly non trivial dynamics emerges as a result of the complex low energy flow of the model.

\textit{The Model, the Local Quantum Critical Point and its variational description - } The model we are going to consider is the so called pseudogap Anderson Impurity Model (pg-AIM) describing a single spin-full fermionic level coupled to a reservoir of non interacting electrons  
\bea
\m{H}_{AIM}=\m{H}_{bath}+\m{H}_{imp}+\m{H}_{hyb}=\sum_{\bk\sigma}\,\eps_\bk\,f^{\dagger}_{\bk\sigma}\,f_{\bk\sigma}+\nonumber\\+\frac{U}{2}\left(n-1\right)^2
+\sum_{\bk\sigma}\,V_\bk\,\left(c^{\dagger}_{\sigma}\,f_{\bk\sigma}+h.c.\right)
\eea
with $n=\sum_{\sigma}\,c^{\dagger}_{\sigma}c_{\sigma}$.
The properties of the bath are encoded in the hybridization function $\Gamma(\eps)=\pi\,\sum_{\bk}\,V_{\bk}^2\,\delta\left(\eps-\eps_{\bk}\right)$, that we assume to vanish as a power law at the Fermi energy $\eps=0$, namely $ \Gamma\left(\eps\right)=
\left(1+r\right)\,\Gamma\,\vert\eps/W\vert^{r}\,\theta(W-\vert\eps\vert)$.
The equilibrium phase diagram of the pg-AIM is extremely rich 	(see Ref.~\cite{Vojta_PhilMag06,Pruschke_RMP} for reviews),  featuring at particle-hole symmetry and for $0<r<1/2$, a quantum phase transition at a critical value of interaction $U_c$ between a Kondo screened phase (for $U<U_c$) and a Local Moment (LM) regime (for $U>U_c$) where the impurity becomes asymptotically free at low energies. As opposite, for $r>1/2$ the only stable fixed point is the local moment one and no Kondo effect can be stabilized for an Anderson Impurity in such a gapless reservoir~\cite{Ingersent_PRB,FritzVojta_prb04,VojtaFritz_prb04}. 
The equilibrium properties of the Kondo to Local Moment quantum criticality have been the subject of a large literature~ (see Ref. \cite{Glossop_prl2011} for a recent work). Quite remarkably though, its out of equilibrium dynamics is an almost unexplored field~\cite{Schiro_long}. 

\begin{figure}[t]
\begin{center}
\epsfig{figure=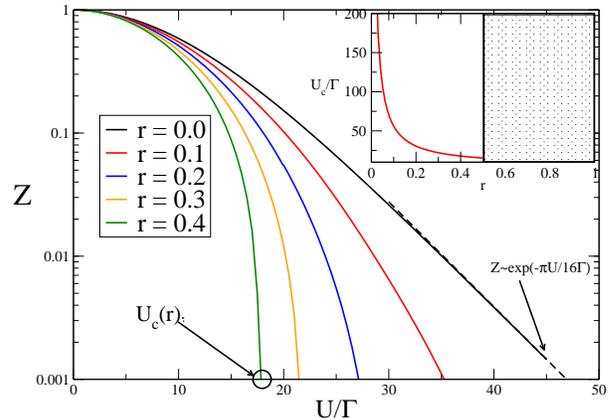,scale=0.325}
\caption{Gutzwiller results for the pg-AIM at half filling. We plot the renormalization factor $Z$ as a function of $U/\Gamma$  for different values of $r$, the power-law exponent of the bath. We see that upon increasing $r$ from the metallic case a critical point emerges which separate a Kondo SC phase ($Z\neq0$) from a LM ($Z=0$) fixed point. Inset: Critical interaction $U_c$ as a function of $r$.} 
\label{fig:aim_qcp}
\end{center}
\end{figure}

As it will be relevant for our study of the dynamics we start our discussion by presenting a simple and intuitive equilibrium variational description of the half filled pg-AIM, in the spirit of the Gutzwiller ansatz, that is able to qualitatively describe salient features of this quantum phase transition. 
We write this correlated ansatz as $\vert\Psi\rangle=\m{P}\,\vert\Phi_{Z}\rangle$
where $\vert\Phi_{Z}\rangle$ is the groundstate of an half-filled non interacting (i.e. $U=0$) resonant level model (RLM) with a variational hybridization $\sqrt{Z}V_k$ while $\m{P}$ acts on the impurity Hilbert space and reads
$\m{P} =\sum_{n=0,1,2}\,\lambda_n\,\vert\,n\rangle\langle\,n\vert$.
Here $\vert\,n\rangle\langle\,n\vert$ projects onto configuration at fixed impurity charge $n$, while $\lambda_n$ are variational parameters.
The key idea behind the above ansatz is that (i) the coupling between the impurity and the reservoir is the crucial quantity to distinguish between the two sides of the QCP and (ii) strong local correlations can substantially renormalize this effective coupling. In order to solve the variational problem we should minimize the energy functional $E = \langle\Psi\,\vert \,H\,\vert\Psi\rangle$ with respect to all the variational parameters . At half filling one can show~\cite{pgaim_si} that only one independent variational parameter enters, that we choose to be $\sqrt{Z}$. This is the so called quasiparticle renormalization factor, which in equilibrium also controls the weight of the low-energy peak in the spectral function.
The saddle point equation reads 
\be\label{eqn:saddle_point}
\frac{U}{8\,\sqrt{1-Z}} +
\frac{2}{\pi\,Z}\,\int\,d\eps\,f(\eps)\,\eps\,A_{Z}\left(\eps\right)=0\,,
\ee
where $A_Z\left(\eps\right)$ is the spectral function of a RLM with a renormalized hybridization $\sqrt{Z}V_k$. Its solution as a function of $U/\Gamma$ and for different bath exponents $r$ is shown in figure~\ref{fig:aim_qcp}.
When $r=0$ (metallic bath) $Z$ smoothly crosses over to an exponentially small value in the strong coupling Kondo regime $U\gg\Gamma$, $Z\sim \exp(-\pi\,U/16\Gamma)$. However as soon as $r\neq0$ it is possible to have a vanishing quasiparticle residue at finite $U$. The critical interaction strength $U_c$ for this to happen can be obtained in closed form $U_c = \frac{16\Gamma}{\pi}\frac{1+r}{r}$
and correctly disappears as $r\rightarrow0$. For $U<U_c$ the variational ground state
has a finite $Z$, meaning the system is a local Fermi Liquid as one expect at the Kondo strong-coupling fixed point. As opposite for $U>U_c$ the impurity is effectively decoupled from the bath, since as $Z\rightarrow0$ also the effective hybridization vanishes. This is a sign the system flows to the LM fixed point. Close to $U_c$ the quasiparticle weight vanishes linearly, $Z\simeq U_c-U$.  
While the above variational analysis provides a simple picture for the quantum phase transition between a Kondo ($Z\neq0$) and a LM ($Z=0$) regime, a more careful treatment of quantum fluctuations would have revealed an even more rich phase diagram. Indeed it has been shown by means of NRG~\cite{Ingersent_PRB} and field theoretical analysis~\cite{FritzVojta_prb04} that such a QCP is present only for $r\in(0,1/2)$, while for $r>1/2$ the system always flow to the local moment fixed point. This is not captured by our simple variational wave function which always shows a finite $U_c$ for all $r>0$. As a result we will confine our analysis to the regime $0<r<1/2$. 
\begin{figure}[t]
\begin{center}
\epsfig{figure=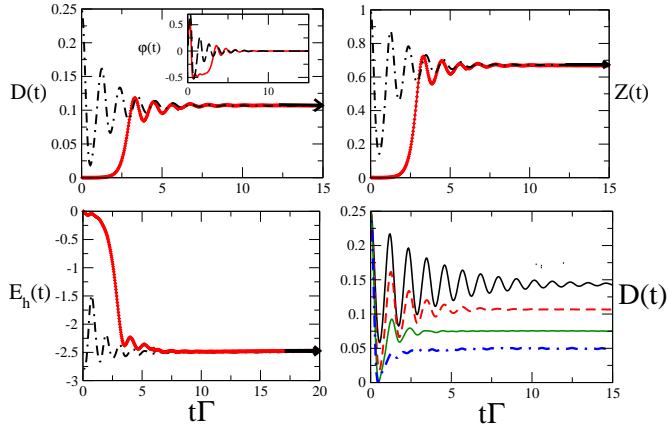,scale=0.325}
\caption{Variational dynamics for the pgAIM with $U/\Gamma=9$, $r=0.4$ after an hybridization (full lines) or  interaction (dashed lines) quench. We plot the impurity double occupancy $D(t)$ and its conjugate variable $\varphi(t)$ (see inset), the renormalized quasiparticle weight $Z(t)$ and the hybridization energy $E_h(t)$. We see that independently from the initial condition the dynamics relaxes to the variational groundstate of the pgAIM model (as shown by the arrows), namely the system thermalizes after a local quench. In the bottom right panel we plot $D(t)$ for different values of $U/\Gamma=5,7,9,11$ from top to bottom. } 
\label{fig:fig2}
\end{center}
\end{figure}

\textit{Dynamics after a local quantum quench - }Let us now consider the dynamics induced in the pgAIM by a \emph{local} quantum quench. In particular we will focus on two classes, namely a sudden switching of (i) the hybridization $V_k$ or (ii) the interaction $U$. Since in a local quantum quench the energy injected into the system does not scale with the size we expect that in the thermodynamic limit, if the system thermalizes, the dynamics will be relaxing toward the ground state of $H_{AIM}$~\cite{Rosch_2011}. In order to study the role of QCP on the dynamics of pgAIM we resort here to a time dependent strongly correlated Gutzwiller variational approach, following recent developments~\cite{SchiroFabrizio_prl10,SchiroFabrizio_PRB11,Lanata_arxiv2011,Lorenzana}. To this extent we introduce an ansatz for time-dependent many body wave function 
\be\label{eqn:tdgutz}
\vert\Psi(t)\rangle = \m{P}(t)\,\vert\Phi(t)\rangle\,,
\ee
where now $\m{P}(t)=\sum_a\,\lambda_n(t)\,e^{i\varphi_n(t)}\,\vert\,n\rangle\langle\,n\vert$ is a time dependent operator acting on the impurity Hilbert space with $\lambda_n,\varphi_n$ being variational parameters, while $\vert\Phi(t)\rangle$ is the solution of a time dependent Schroedinger equation with renormalized non interacting hamiltonian whose form will be specified in the following.
\begin{figure}[t]
\begin{center}
\epsfig{figure=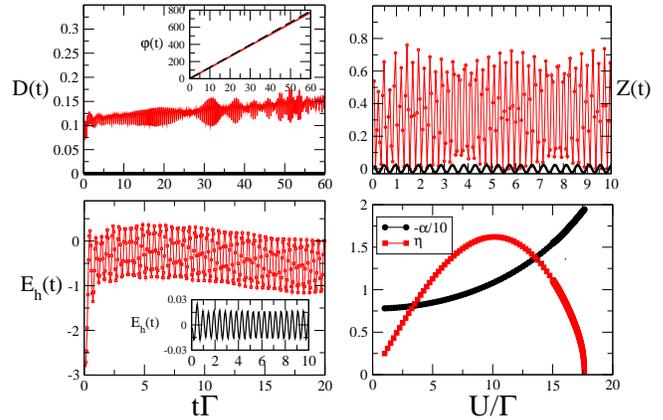,scale=0.325}
\caption{Variational dynamics for the pgAIM ($r=0.4$) after a local quench across the SC-LM quantum critical point. Upper panel: dynamics of double occupancy $D(t)$ and quasiparticle weight $Z(t)$ for an interaction (red) or hybridization (black) quench, with $U/\Gamma=25$. In both cases local quantities do not relax. Bottom panels: the average hybridization between the impurity and the bath shows coherent undamped oscillations (left) for both type of quenches. This can be traced back to the vanishing at $U_c$ of the energy scale $\eta$ controlling the coupling between the impurity sector and the bath modes (see right and main text).} 
\label{fig:fig3}
\end{center}
\end{figure}
 The variational dynamics is then obtained by making the following real time action 
\be
\m{S}=\int\,dt\,\langle\Psi(t)\vert i\partial_t-H\vert\Psi(t)\rangle
\ee
stationary with respect to all the variational parameters $\lambda_n,\varphi_n,\vert\Phi(t)\rangle$.
We quote the final result of this calculation, which goes along the same line of previous case~\cite{pgaim_si}. Assuming to start from a particle-hole initial preparation for the dot and the reservoir, then the dynamics for double occupancy $D(t)$ and its conjugate variable $\varphi(t)$ reads
\bea\label{eqn:var_dyn1}
\dot{\varphi} =4\,E_{hyb}(t)\,\sqrt{D\left(1/2-D\right)}\,sin\varphi \\
\label{eqn:var_dyn2}
\dot{D} =U+E_{hyb}(t)\,\frac{1-4\,D}{\sqrt{D\left(1/2-D\right)}}\,cos\varphi
\eea
with $E_{hyb}(t)=\langle\Phi(t)\vert\,H_{hyb}\vert\Phi(t)\rangle$, while the state $\vert\Phi(t)\rangle$ satisfies the Schroedinger equation of a RLM
\be\label{eqn:var_dyn3}
i\partial_t\,\vert\Phi(t)\rangle = \left(H_{bath}+\sqrt{Z(t)}\,H_{hyb}\right)
\vert\,\Phi(t)\rangle
\ee
with a time-dependent hybridization 
$
\sqrt{Z(t)}=4\,\sqrt{D(t)\,\left(1/2-D(t)\right)}\,cos\varphi(t)\,.
$
It is important to stress at this point that the above dynamics fully accounts for the coupling between the variational parameters $D,\varphi$ entering the projector and the fermionic degrees of freedom in the wave function $\vert\Phi(t)\rangle$. Let us now discuss the solution of the variational dynamics (\ref{eqn:var_dyn1}-\ref{eqn:var_dyn3}) by firstly considering local quenches that end in the Kondo screened phase, namely for $U<U_c$. In figure~\ref{fig:fig2}  we plot, for the two kind of local quenches we are interested and for a bath exponent $r=0.4$, the dynamics of three relevant quantities namely the double occupancy $D(t)$, its conjugate phase $\varphi(t)$ and the hybridization energy $E_{hyb}(t)$. We see that after an initial transient, for time scales larger than some relaxation time $\tau_{rel}$ the dynamics reaches a steady state which is independent from the initial condition. A direct comparison shows that such a steady state coincide with the (variational) ground state of the pseudo-gap AIM, that we plot as a dashed line in the figure. In other words the system thermalizes after a local quench. However, a very different dynamical behaviour emerges when the system is quenched deep in the LM regime, as we show in figure~\ref{fig:fig3}. In particular the double occupancy shows very fast oscillations around an average value which is different from the expected ground state value. More interestingly, the hybridization strength, which measure the effective coupling between the impurity and the bath, starts oscillating around zero. This enlights the fact that above the critical point, as the impurity effectively decouples from the bath, the dynamics cannot equilibrate.

\emph{Discussion - }
The thermalization of the pgAIM for $U<U_c$ is consistent with the behaviour of the conventional AIM in a metallic bath~\cite{Nordlander_99,Kehrein_Lobaskin_05,tnrg_Anders,Anders_prb06}. It reflects the fact that, as the low-energy/long-time limit is approached, the model flows toward the SC fixed point where the impurity stays always coupled to the reservoir and the system behaves as a local Fermi Liquid~\cite{Nozieres_FL}. This phase is adiabatically connected to the weak coupling RLM whose dynamics after a local quench leads to thermalization. A closer look to the variational dynamics gives further insights into this thermalization process. Indeed one can explicitly show~\cite{pgaim_si} that the 
equilibrium ground-state variational solution is a steady state (SS) of the dynamics~(\ref{eqn:var_dyn1},\ref{eqn:var_dyn2},\ref{eqn:var_dyn3}). This is immediately seen for the classical degrees of freedom by noticing that (i) $\dot{\varphi}=0$ implies $sin\,\varphi_{\star}=0$ in the SS and (ii) after simple algebra the equation~(\ref{eqn:var_dyn2}) becomes the saddle point condition~(\ref{eqn:saddle_point}) for $Z_{\star}$, provided the hybridization energy of the RLM thermalizes at long time, $E_{hyb}(t\gg\tau_{rel})=E^{\star}_{hyb}$, as one can show to be the case~\cite{pgaim_si} . This result suggests to linearize the dynamics around the SS and to extract information on the relaxation time upon approaching the QCP. A simple but tedious calculation gives the following reduced dynamics~\cite{pgaim_si} for the double occupation $\delta D(t)=D(t)-D_{\star}$
\be\label{eqn:linear}
\delta \ddot{D}  = \alpha\,\delta D +\eta\,\delta E_{hyb}
\ee
in terms of $\delta E_{hyb}=E_{hyb}(t)-E_{hyb}^{\star}$, the deviation from steady state of the average hybridization.  The coefficients $\alpha$ and $\eta$ give information on characteristic relaxation timescales and are purely equilibrium steady state properties. They read respectively $\alpha\sim\,E^{\star2}_{hyb}/Z_{\star}$ and $\eta\sim\,(1-4D_{\star})\,E^{\star}_{hyb}$.
Their behaviour with $U$, as plotted in figure~\ref{fig:fig3}, reveal that while $\alpha$ is finite (and negative) approaching the QCP as $E^{\star}_{hyb}\sim\sqrt{Z_{\star}}$, the coupling $\eta$ between the impurity sector and the bath modes vanishes at $U_c$ as $\eta\sim\sqrt{Z_{\star}}\sim~\sqrt{U_c-U}$. 
This vanishing energy scales suggests a diverging time-scale for thermalization (and relaxation) of local observables. As a result for quenches above the QCP no steady state can be obtained within the above variational dynamics, as already revealed by the numerical solution plotted in figure \ref{fig:fig3}.

It is interesting to discuss whether the above picture is robust against quantum fluctuations that are not included in the Gutzwiller variational scheme. We expect these to play a special role on the LM side of the QCP, to damp out the oscillations and eventually lead to a (non-thermal) steady state. In addition we expect that around $U_c$, and particularly for quenches ending at the QCP, important effects beyond mean field will be crucial to capture correctly the quantum critical relaxation dynamics.  A possible framework to treat non-equilibrium effects beyond this time dependent variational scheme is offered by the slave spin formulation of the AIM~\cite{SchiroFabrizio_PRB11,Baruselli_Fabrizio_PRB12}, where the impurity electron is represented as an auxiliary fermion coupled to a two-level system (TLS). Quite interestingly , the Kondo effect emerges in this picture as a localized phase of an effective sub-ohmic spin boson model~\cite{Baruselli_Fabrizio_PRB12}. As far as the non equilibrium dynamics is concerned, while a simple mean field decoupling of the TLS and fermionic sectors leads to an evolution which is fully equivalent to the time-dependent Gutzwiller discussed in this paper, quantum fluctuations beyond  the saddle point can be included systematically. These result into coupled dynamical equations for the auxiliary fermionic and TLS self-energies, similar to large N approaches which are known to capture some aspects of the quantum critical relaxation dynamics~\cite{Vojta_PhilMag06,Vojta_LargeN_prl01}. These issues represent interesting directions for further investigations. In spite of that, we could say that the qualitative picture of the non equilibrium dynamics, including a diverging time scale and the absence of thermalization for quenches in the LM regime appear as a robust features of the pgAIM model that are well captured by our simple variational wave function. In this respect it is interesting to notice that a recent investigation of the nonequilibrium dynamics in the Ferromagnetic (FM) Kondo model~\cite{KondoFerro,Hackl_PRB2009}, that always flows to the LM fixed point,  finds a similar lack of ergodicity in the long time dynamics.

Finally, we conclude by discussing possible connections with experiments. As we mentioned in the introduction, recently the non-equilibrium dynamics of an AIM after a local quench has been addressed, indirectly, in experiments on single self-assembled quantum dots~\cite{Imamoglu_nature11} by measuring the absorption/emission spectrum after optical excitations. This spectrum turns to be related to the statistics of the work done during the quench~\cite{Tureci_prl11,Heyl_Kehrein_arxiv2010}, which features a clear signature of the Kondo effect in the form of an edge singularity at low values of the work. A natural question would be therefore whether one can arrange a system with a similar Kondo to Local Moment QCP in these systems and how to characterize this quantum phase transition from the point of view of the non-equilibrium dynamics after a local quench. As far as the first point is concerned, one obvious possibility would be to play with the conduction electrons density of states, so to arrange for a semi-metallic behaviour. Since the conduction electrons are provided by a doped semiconductor (typically n-doped GaAs), this may be hard to realize in practice. A more interesting scenario would be to probe the dynamics of a system of two Kondo spins anti-ferromagnetically coupled, which is known to posses a similar quantum phase transition~\cite{JonesVarma_prb89}. In that case, a clear dynamical signature of the QCP would emerge by looking at the asymptotic behaviour of the optical spectrum, where the X-ray edge singularity is expected to be cut-off for a much weaker (logarithmic) singularity.

\textit{Conclusions -} 
In this paper we have discussed the nonequilibrium dynamics after a local quench in the pgAIM, as a prototype of fermionic impurity quantum critical point. We have argued that equilibration in this problem is deeply connected to the occurrence of Kondo Effect at low energies. While for local quenches in the Kondo phase thermalization occurs, this is not the case when quenching across the QCP deep in the LM regime.  This is correctly captured by a simple and intuitive time dependent variational wave function which in addition is able to describe the divergence of the relaxation time as the QCP is approached. 

\textit{Acknowledgement - } It is a pleasure to thank N. Andrei, G. Biroli, M. Fabrizio,  D. Huse,  J. Macjeiko, H. Tureci and M. Vojta for interesting discussions and comments related to this work.

\bibliographystyle{apsrev}


\subsection{Gutzwiller wave function for the AIM}

Here we briefly sketch the main results of the Gutzwiller variational approach to the AIM~\cite{Schonhammer_prb76,Schonhammer_prb90,Fabrizio_Vietri,Fabrizioprb07}. We write the hamiltonian as
\bea
\m{H}_{AIM}=\m{H}_{bath}+\m{H}_{imp}+\m{H}_{hyb}=\sum_{\bk\sigma}\,\eps_\bk\,f^{\dagger}_{\bk\sigma}\,f_{\bk\sigma}+\nonumber\\+\frac{U}{2}\left(n-1\right)^2
+\sum_{\bk\sigma}\,V_\bk\,\left(c^{\dagger}_{\sigma}\,f_{\bk\sigma}+h.c.\right)
\eea
with $n=\sum_{\sigma}\,d^{\dagger}_{\sigma}d_{\sigma}$ and consider a wave function of the form
\be 
\vert\Psi\rangle=\m{P}\,\vert\Phi_{\lambda}\rangle
\ee
where $\vert\Phi_{\lambda}\rangle$ is the groundstate of a non interacting (i.e. $U=0$)  half filled resonant level model (RLM) with a variational hybridization $\lambda\,V_k$ while $\m{P}$ acts on the impurity Hilbert space and reads $\m{P} =\sum_{n=0,1,2}\,\lambda_n\,\vert\,n\rangle\langle\,n\vert$.
Here $\vert\,n\rangle\langle\,n\vert$ projects onto configuration at fixed impurity charge $n$, while $\lambda_n$ are variational parameters, related to the correlated probability of having a given charge $n$ on the impurity. This latter quantity is defined as
\be 
P_n = \langle\Psi\vert\,n\rangle\langle\,n\vert\Psi\rangle=\lambda_n^2\,\langle\Phi_{\lambda}\vert\,n\rangle\langle\,n\vert\Phi_{\lambda}\rangle
\ee
We can impose without loss of generality the normalization condition on the wave function
\be
\langle\Psi\vert\Psi\rangle = \langle\Phi_{\lambda}\vert\,\m{P}^2\,\vert\Phi_{\lambda}\rangle=1\,,
\ee
that also implies $\sum_n\,P_n=1$. In addition one can easily convince that for the above wave function the following relation holds at particle-hole (PH) symmetry
\be\label{eqn:gutzw1}
\langle\Phi_{\lambda}\vert\,\m{P}^2\,\hat{n}\vert\Phi_{\lambda}\rangle=
\langle\Phi_{\lambda}\vert\,\hat{n}\vert\Phi_{\lambda}\rangle\,=1,
\ee
or equivalently $\sum_n\,n\,P_n=1$. Since at PH symmetry we have $P_0=P_2\equiv\,D$ from these two latter conditions we conlcude also that $P_1=1-2D$.
In order to solve the variational problem we should minimize the energy functional $E_{var}= \langle\Psi\,\vert \,H\,\vert\Psi\rangle$ with respect to all the variational parameters. The identity~(\ref{eqn:gutzw1}) greatly simplifies the evaluation of the variational energy. Indeed using this result one can show that~\cite{Fabrizioprb07}
\be
\langle\Psi\vert\,H_{bath}\vert\Psi\rangle =  
\langle\Phi_{\lambda}\,\vert\,H_{bath}\vert\Phi_{\lambda}\rangle\qquad
\ee
as well as
\be 
\langle\Psi\vert\,H_{hyb}\vert\Psi\rangle =  
\sqrt{Z}\langle\Phi_{\lambda}\,\vert\,H_{hyb}\vert\Phi_{\lambda}\rangle\qquad\,
\ee
where $\sqrt{Z}$ is a function of $P_n$ that at half filling reads $\sqrt{Z}=4\sqrt{D\left(1/2-D\right)}$. 
Using these results we can write the variational energy as
\bea
E_{var}=  \langle\Psi\,\vert \,H\,\vert\Psi\rangle=
\langle\Phi_{\lambda}\,\vert\,H_{bath}\vert\Phi_{\lambda}\rangle+\nonumber\\
+\sqrt{Z}\langle\Phi_{\lambda}\,\vert\,H_{hyb}\vert\Phi_{\lambda}\rangle+UD
\eea
Variation with respect to $\vert\Phi_{\lambda}\rangle$ at fixed $D$ gives a renormalized RLM hamiltonian with $\lambda=\sqrt{Z}$. The saddle point equation for $Z$ is then obtained by taking the derivative of the variational energy. Using the fact that for a RLM with hybridization $\sqrt{Z}\,V_k$ we have $\partial\,E_{gs}/\partial\sqrt{Z}= \langle\,H_{hyb}\rangle_{gs}$
and by writing this latter average as
\be
\langle\,H_{hyb}\rangle_{gs} = \frac{4}{\pi\,\sqrt{Z}}\,\int\,d\eps\,f(\eps)\,\eps\,A_{Z}\left(\eps\right)
\ee
we obtain the equation in the main text,
where we have defined $A_Z\left(\eps\right)$ as the spectral function of a RLM with a renormalized hybridization $\sqrt{Z}V_k$
\be
A_{Z}\left(\eps\right) = \frac{Z\,\Gamma\left(\eps\right)}
{\left(\eps-Z\,\Delta_R(\eps)\right)^2+\left(Z\,\Gamma(\eps)\right)^2}\,,
\ee
with
\be
\Delta_R(\eps) = \mbox{P}\int\,\frac{d\omega}{\pi}\,\frac{\Gamma(\omega)}{\eps-\omega}\,,\qquad
\Gamma(\eps)=\pi\,\sum_k\,V_k^2\,\delta(\eps-\eps_k)\,.
\ee
\section{Time Dependent Gutzwiller for the AIM}

Extensions of the Gutzwiller wave function to the time dependent case have been recently introduced, both for lattice models~\cite{SchiroFabrizio_prl10} and for quantum impurities~\cite{Lanata_arxiv2011}. The evaluation of the real-time action on the time dependent wave function
\be
\vert\Psi(t)\rangle = \m{P}(t)\,\vert\Phi(t)\rangle\,,
\ee
where now $\m{P}(t)=\sum_a\,\lambda_n(t)\,e^{i\varphi_n(t)}\,\vert\,n\rangle\langle\,n\vert$
while $\vert\Phi(t)\rangle$ is the time dependent wave function of a proper non interacting RLM,  can be done along similar lines as in the equilibrium case. After imposing the normalization condition 
\be
\langle\Phi(t)\vert\,\m{P}^2(t)\,\vert\Phi(t)\rangle =1\,,\qquad
\ee
and noticing that at PH simmetry we have
\be
\langle\Phi(t)\vert\,\m{P}^2(t)\,n\,\vert\Phi(t)\rangle =
\langle\Phi(t)\vert\,\,n\,\vert\Phi(t)\rangle=
1
\ee
we obtain the following real-time action
\bea\label{eqn:action}
\m{S}=\int\,dt\,\langle\Psi(t)\vert i\partial_t-H\vert\Psi(t)\rangle=\nonumber\\
  =\m{S}_{RLM}+\int\,dt\,
\left(i\dot{\varphi}\,D-U\,D\right)
\eea
where $\m{S}_{RLM}$ is the action of an effective time dependent RLM that is
\be
 \m{S}_{RLM} = \int\,dt\,\langle\Phi(t)\vert\,i\partial_t-H_{RLM}(t)\vert\Phi(t)\rangle\,.
\ee
with hamiltonian $H_{RLM}(t)=H_{bath}+\sqrt{Z(t)}\,H_{hyb}$ where the time dependent renormalization factor $\sqrt{Z}$ reads
\be
\sqrt{Z}=4\sqrt{D\left(1/2-D\right)} \,cos\varphi
\ee

From the action~(\ref{eqn:action}) we get the variational dynamics for $D(t)$ and $\varphi(t)$ 
\bea\label{eqn:var1}
\dot{D} = 4\,E_{hyb}(t)\,\sqrt{D\left(1/2-D\right)}\,sin\varphi\\
\label{eqn:var2}
\dot{\varphi} = U +E_{hyb}(t)\,\frac{1-4D}{\sqrt{D\left(1/2-D\right)}}\,cos\varphi
\eea
and the Schoedinger equation for $\vert\Phi(t)\rangle$
\be
i\partial_t\,\vert\Phi(t)\rangle =H_{RLM}(t)
\vert\,\Phi(t)\rangle\,.
\ee
In order to solve the
dynamics of the effective time dependent RLM we can proceed with an equation of 
motion appoach as in ~\cite{Lanata_arxiv2011}. Let us define the following averages
\be
W_{k} = \sum_{\sigma}\langle\,f^{\dagger}_{k\sigma}\,c_{\sigma}\rangle\,,\qquad
N_{kp} = \sum_{\sigma}\langle\,f^{\dagger}_{k\sigma}\,f_{p\sigma}\rangle\,
\ee
whose dynamics is obtained by evaluating the commutators with $H_{RLM}(t)$
\bea\label{eqn:var3}
i\,\dot{W}_{k} =-\eps_k\,W_{k} + \sum_{p}\,\bar{V}_p\,
\,N_{kp}-\bar{V}_k\,n\\
\label{eqn:var4}
i\,\dot{N}_{kp} =
\left(\eps_p-\eps_k\right)\,N_{kp} +
\bar{V}_p\,W_{k} - \bar{V}_k\,W_{p}^{\star}\,\\
\label{eqn:var5}
i\,\dot{n} = -
\sum_p\,\bar{V}_p\,\left(W_{p}-W_{p}^{\star}\right)\,.
\eea
where we have defined $\bar{V}_k(t)=\sqrt{Z(t)}\,V_k$. We note that, since PH symmetry is preserved at any time even for a time dependent hybridization, the occupation of the dot is fixed and the last equation is irrelevant, as $n=1$.
The hybridization strength entering the classical equations of motion read
\be
E_{hyb}(t) = \sum_{k}\,V_k\,\left(W_{k}+h.c.\right)
\ee

\subsection{Long time dynamics and Thermalization}

We now want to show that the equilibrium groundstate variational solution is a fixed point of the above dynamics. In order to see this let us start from the classical dynamics and search for a stationary solution, $D_{\star},\varphi_{\star}$. By putting derivatives to zero we get
\be
sin\varphi_{\star}=0 
\ee
and
\be\label{eqn:ss}
0 = U +E^{\star}_{hyb}\,\frac{1-4D_{\star}}{\sqrt{D_{\star}\left(1/2-D_{\star}\right)}}
\ee
where we have defined
\be
E^{\star}_{hyb}= \mbox{lim}_{t\rightarrow\infty}\,E_{hyb}(t)
\ee
as the steady state limit of the hybridization energy. Equation~(\ref{eqn:ss}) is nothing but the equilibrium saddle point energy condition for $D$
\bea
\frac{\partial E^{gs}}{\partial D} = U+\frac{\partial E^{gs}_{hyb}}{\partial D} = U+ 
\frac{\partial E^{gs}_{hyb}}{\partial \sqrt{Z}}\,\frac{\partial\sqrt{Z}}{\partial D}=\\
=U + E^{gs}_{hyb}\,\frac{1-4D}{\sqrt{D\left(1/2-D\right)}}=0
\eea
provided that the steady state value of the hybridization energy, $E^{\star}_{hyb}$, coincides with the corresponding ground state value $E^{gs}_{hyb}$ as a function of $\sqrt{Z_{\star}}$, namely
\be 
E^{\star}_{hyb}(\sqrt{Z_{\star}} ) = E^{gs}_{hyb}(\sqrt{Z_{\star}})\,.
\ee
Therefore in order to proceed further we have to show that $E^{gs}_{hyb}$ is the steady state 
of the variational dynamics for the bath degrees of freedom, or in other words that the auxiliary RLM thermalizes to the groundstate with hybridization $\sqrt{Z_{\star}}\,V_k$.

This can be seen explicitly by looking at the dynamics of the bath degrees of freedom. To this extent we plug on the right hand side of equations~(\ref{eqn:var3},\ref{eqn:var4}) the equilibrium ground-state values of $W_k$ and $N_{kp}$ that for a RLM with hybridization $\sqrt{Z_{\star}}\,V_k$ can be obtained by standard methods~\cite{Mahan_book}. These read
\be
W^{gs}_{k}=T\,\sum_{i\omega}\,e^{i\omega 0^+}\,
\m{G}^{\star}(i\omega)\,\frac{\bar{V}_k}{i\omega-\eps_k}
\ee
as well as
\be
N^{gs}_{kp}=\delta_{kp}f(\eps_k)+
T\,\sum_{i\omega}\,e^{i\omega\,0^+}\,
\frac{\bar{V}_p}{i\omega-\eps_p}\,\m{G}^{\star}(i\omega)\,
\frac{\bar{V}_k}{i\omega-\eps_k}
\ee
with $\bar{V}_p=\sqrt{Z_{\star}}\,V_p$, the impurity Green's function $\m{G}^{\star}(i\omega)$ given by
\be
 \m{G}^{\star}(i\omega)=\frac{1}{i\omega-Z_{\star}\,\Delta(i\omega)}
\ee
and the hybridization function 
$$
\Delta(i\omega)=\sum_k\,\frac{V_k^2}{i\omega-\eps_k}
$$
If we plug these expressions into the variational equations of motion we can see they represent stationary solutions. We find indeed
\bea
 i\,\dot{N}_{kp} &=&
\left(\eps_p-\eps_k\right)\,T\,\sum_{i\omega}\,e^{i\omega\,0^+}\,
\frac{\bar{V}_p\,\m{G}^{\star}(i\omega)\,\bar{V}_k}{\left(i\omega-\eps_p\right)\left(i\omega-\eps_k\right)}+\nonumber\\
&&+T\,\sum_{i\omega}\,e^{i\omega 0^+}\,
\m{G}^{\star}(i\omega)\,\left(\frac{\bar{V}_p\,\bar{V}_k}{i\omega-\eps_k}-
\frac{\bar{V}_k\,\bar{V}_p}{i\omega-\eps_p}
\right)=\nonumber\\
&=&0
\eea
as well as
\bea
i\,\dot{W}_k &=&  -T\,\sum_{i\omega}\,e^{i\omega0^+}\,\m{G}^{\star}(i\omega)\,\frac{\eps_k\,\bar{V}_k}{i\omega-\eps_k}+\bar{V}_k\left(f(\eps_k)-1\right)+\nonumber\\
&&+T\,\sum_{i\omega}\,e^{i\omega0^+}\,\Delta^{\star}(i\omega)\,\m{G}^{\star}(i\omega)\,
\frac{\bar{V}_k}{i\omega-\eps_k}=0
\eea
where in the last step we have used the fact that
\be
 \Delta^{\star}(i\omega)\,\m{G}^{\star}(i\omega)=-1+i\omega\,\m{G}^{\star}(i\omega)\,.
\ee
This shows in particular that the groundstate value of the hybridization $E^{gs}_{hyb}$, which immediately follows from $W_k^{gs}$, is the steady state of the variational dynamics.

\subsubsection{Linearization around the steady state}

In this section we linearize the variational dynamics around the steady state. To this extent it is convenient to write the equations of motion in a different form. Indeed the average hybridization 
$E_h$ is written as
\be
 E_{hyb} = \,\sum_k\,V_k\,\left(W_k+h.c.\right)=
 2\,\sum_k\,V_k\,W_k^\mR
\ee
which suggests to express the dynamics in terms of the real and imaginary parts, $W_k^{R,I}$ and $N_{kp}^{\mR,\mI}$. A simple algebra gives
\bea\label{eqn:bath}
\dot{W}_k^\mR = -\eps_k\,W_k^\mI+\sum_p\,\bar{V}_p\,N_{kp}^\mI\\
\dot{W}_k^\mI = \eps_k\,W_k^\mR-\sum_p\,\bar{V}_p\,N_{kp}^\mR+\bar{V}_k\\
\dot{N}_{kp}^{\mR} = \left(\eps_p-\eps_k\right)\,N_{kp}^\mI+
\bar{V}_p\,W_k^\mI+\bar{V}_k\,W_p^\mI\\
\dot{N}_{kp}^{\mI} = -\left(\eps_p-\eps_k\right)\,N_{kp}^\mR-
\bar{V}_p\,W_k^\mR-\bar{V}_k\,W_p^\mR
\eea
These equations, together with the dynamics fo $D$ and $\varphi$
\bea
\dot{D} = \,\sqrt{Z}\,E_{hyb}(t)\,sin\varphi\\
\dot{\varphi} = U +E_{hyb}(t)\,\frac{\partial\sqrt{Z}}{\partial\,D}\,cos\varphi
\eea
with $\sqrt{Z}=4\sqrt{D(1/2-D)}$, represents the variational dynamics for the variables $\left(D,\phi,W_k^\mR,W_k^\mI,N_{kp}^\mR,N_{kp}^\mI\right)$
that we want to linearize around the steady state values, $\left(D_{\star},\phi_{\star},W_{k\star}^\mR,W_{k\star}^\mI,N_{kp\star}^\mR,N_{kp\star}^{\mI}\right)$.

It is a straightforward but tedious calculation to linearize this dynamics and obtain a system of first order linear differential equations. In principle, by computing the eigenmodes of the associated matrix we could get information of the stability of the steady state solution. However in order to make analytical progress it is useful to express the dynamics as a second order differential equation for half of the variables. This dynamics reads
\bea
\delta \ddot{D}  = \alpha\,\delta D +\sum_k\,\eta_k\,\delta W_k^{\mR}\\
\delta \ddot{W}_k^{\mR}  =  \sum_q\,\Lambda_{kq}\,\delta\,W_q^{\mR}+
\sum_q\,\Gamma_q\,\delta N_{kq}^{\mR}\,+\beta_k\,\delta D\\
\delta \ddot{N}_{kp}^{\mR}  =  \sum_{qq'}\,\Omega^{kp}_{qq'}\,\delta\,N_{q'q}^{\mR}+
\sum_q\,\Theta^{kp}_q\,\delta W_{q}^{\mR}\,+\gamma_{kp}\,\delta D
\eea
where the coefficients $\alpha,\eta_k,\Lambda_{kq},\Gamma_q,\beta_k,\Omega_{qq'}^{kp},\Theta_{q}^{kp},\gamma_{kp}$
read respectively
\bea
\alpha &=& \sqrt{Z_{\star}}\,(E^{\star}_{hyb})^2\,\frac{\partial^2\sqrt{Z_{\star}}}{\partial\,D^2}\,cos^2\varphi_{\star}\\
\eta_k &=& 2\,\sqrt{Z_{\star}}\,\frac{\partial\sqrt{Z_{\star}}}{\partial D}\,cos^2\varphi_{\star}\,V_k\,E^{\star}_{hyb}\\
\beta_k &=& \eps_k\,\frac{\partial\sqrt{Z_{\star}}}{\partial D}\,cos\varphi_{\star}\,\sum_p\,V_p\,N_{kp\star}^\mR+\nonumber\\
&-&\left(\sqrt{Z_{\star}}\,\frac{\partial\sqrt{Z_{\star}}}{\partial\,D}\right)\,
\sum_q\left(V_q^2\,W_{k\star}^{\mR}+V_qV_k\,W_{q\star}^{\mR}\right)\\
\gamma_{kp}&=&\left(\eps_k-\eps_p\right)
\left(\frac{\partial\sqrt{Z_{\star}}}{\partial D}\,\left(V_p\,W_{k\star}^{\mR}+
V_k\,W_{p\star}^{\mR}\right)\right)+\nonumber\\
&-&\sqrt{Z_{\star}}\,\frac{\partial\sqrt{Z_{\star}}}{\partial D}\,
\sum_q\left(V_qV_p\,N_{kp\star}^\mR+V_qV_k\,N_{pq\star}^\mR
\right)\\
\Gamma_q &=& -\sqrt{Z_{\star}}\,cos\varphi_{\star}\,V_q\,\eps_q\\
\Lambda_{kq} & = &-
\delta_{kq}\left(\eps_q^2+Z_{\star}\,cos^2\varphi_{\star}\,V_q\right)+\nonumber\\
&-&Z_{\star}\,cos^2\varphi\,V_q\,V_k\\
\Omega_{qq'}^{kp}&=&-\left(\eps_q-\eps_{q'}\right)^2\,\delta_{qp}\,\delta_{q'k}+\\
&+& Z_{\star}\,cos^2\varphi_{\star}\,\left(V_pV_q\delta_{q'k}+V_k\,V_q\,\delta_{q'p}\right)\\
\Theta_q^{kp}&=&\sqrt{Z_{\star}}\,cos\varphi_{\star}\,
\left(
\delta_{kp}\,V_p\,\left(2\eps_q-\eps_p\right)+\delta_{pq}\,V_k\eps_k
\right)
\eea

\bibliographystyle{apsrev}

\end{document}